\documentclass[runningheads]{llncs}
\usepackage{doc}
\pdfoutput=1
\usepackage{amssymb,amsmath,amsfonts,bbm}
\usepackage{caption,subcaption,multirow}
\usepackage{xcolor,graphicx,listings}
\usepackage[linesnumbered,ruled]{algorithm2e}
\usepackage{url}

\newcommand{\N}{\mathbf{\mathcal{N}}}
\newcommand{\I}{\mathbf{\mathcal{I}}}
\newcommand{\f}{\mathbf{f}}

\definecolor{mygreen}{rgb}{0,0.6,0}
\definecolor{mygray}{rgb}{0.5,0.5,0.5}
\definecolor{mymauve}{rgb}{0.58,0,0.82}

\title{Computing Derivatives for PETSc Adjoint Solvers using Algorithmic Differentiation}

\titlerunning{Algorithmic Differentiation for PETSc Adjoint Solvers}

\author{Joseph G. Wallwork\inst{2}\fnmsep
\thanks{The present work was done during an internship at Argonne National Laboratory.}
\and Paul D. Hovland\inst{1}
\and Hong Zhang\inst{1}
\and Oana Marin\inst{1}}

\authorrunning{J. G. Wallwork, P. Hovland, H. Zhang and O. Marin}
\institute{Argonne National Laboratory, 9700 Cass Avenue, Lemont, IL, 60439, U.S.A.
\and
Imperial College London, Kensington, London, SW7 2AZ, U.K.\\
\email{j.wallwork16@imperial.ac.uk}\\
}

\begin{document}
\maketitle

\begin{abstract}
Most nonlinear partial differential equation (PDE) solvers require the Jacobian matrix associated to the differential operator.
In PETSc \cite{petsc-user-ref}, this is typically achieved by either an analytic derivation or numerical approximation method such as finite differences.
For complex applications, hand-coding the Jacobian can be time-consuming and error-prone, yet computationally efficient.
Whilst finite difference approximations are straight-forward to implement, they have high arithmetic complexity and low accuracy.
Alternatively, one may compute Jacobians using \emph{algorithmic differentiation (AD)}, yielding the same derivatives as an analytic derivation, with the added benefit that the implementation is problem independent.
In this work, the \emph{operator overloading} AD tool ADOL-C \cite{ADOLC} is applied to generate Jacobians for time-dependent, nonlinear PDEs and their adjoints.
Various strategies are considered, including compressed and matrix-free approaches.
In numerical experiments with a 2D diffusion-reaction  model, the performance of these strategies has been studied and compared to the hand-derived version.

\end{abstract}

\section{Introduction}\label{sec:Intro}

The adjoint method has proved to be an indispensable tool in computational modelling and optimization, enabling sensitivity analysis, goal-oriented error estimation and data assimilation, for example (see \cite{Gunzburger2003}).
It is used to compute the derivative of an objective function with respect to parameters of interest, with a cost independent of the number of parameters.
Algorithmic differentiation (AD), as a traditional engineering approach, has been developed to automatically produce an adjoint code that computes the derivatives. 
AD tools take as input a forward model that users implement in a low-level language, and derive the adjoint model in a line-by-line fashion, through source-to-source transformations, operator overloading or a combination thereof.
While this black-box approach gives the highest degree of automation and requires least knowledge of the mathematical models, it suffers from many low-level implementation-specific difficulties such as memory allocation, pointers, I/O and parallel communication (e.g. MPI and OpenMP).
Recently new high-level AD tools/libraries such as \texttt{libadjoint} \cite{Farrell2013}, \texttt{FATODE} \cite{Zhang2014} and the \texttt{TSAdjoint} component in PETSc \cite{Abhyankar2018,Zhang2017} have been developed to operate on high-level systems; therefore concerns on low-level implementations can be avoided.

In the high-level adjoint approach, one has to solve an adjoint equation that is derived either before discretization or after discretization, corresponding to the continuous adjoint method and the discrete adjoint method, respectively.
The derivation is fully automated in \texttt{libadjoint} if one specifies the forward model in a high-level language similar to mathematical notation.  
In contrast, \texttt{FATODE} and PETSc \texttt{TSAdjoint} provide a built-in implementation of the discrete adjoint model derived based on the time stepping algorithms for solving ordinary differential equations (ODEs); simulation of time-dependent PDEs is abstracted as a sequence of time steps and the libraries differentiate each time step essentially.

Nevertheless, the core task of these high-level AD tools is to assemble user-supplied callback functions that calculate function derivatives (a.k.a. Jacobians).
The Jacobian functions may be written manually,
% or may be
approximated by finite differences, or
% may be
generated with an AD tool.

Writing an analytical Jacobian matrix evaluation function is an error-prone, tedious and time-consuming task.
PETSc offers tools to calculate a finite difference approximation of the Jacobian matrix suitable for some classes of problems.
With the aim of developing a framework for performing automatic and efficient large scale adjoint calculation, the present work extends PETSc's functionality to automate the process of Jacobian computation, by applying ADOL-C \cite{ADOLC} within the higher level time stepping solver.
Although we focused on the discrete adjoint, the Jacobian automation is also applicable and useful for the continuous adjoint approach.

Previous efforts have been made to use source-to-source transformation AD for the numerical solution of PDEs using PETSc (see \cite{Hovland2005}).
The present work goes further - with a focus on adjoint problems - and computes the required Jacobian transpose.
Futhermore, we exploit the reverse mode of AD to compute the \emph{matrix-free} Jacobian transpose vector product.
% Another difference is that the present work uses \emph{operator overloading} AD, rather than source-to-source transformation.

The paper is organized as follows. We begin by introducing the AD tool ADOL-C.
In Section \ref{sec:PETScAdjoint}, we briefly introduce the PETSc adjoint solver.
In Section \ref{sec:example}, we describe an exemplar adjoint problem based on a nonlinear PDE.
Section \ref{sec:OutlineAndFeatures} presents our strategies to automate the Jacobian computation in PETSc. Numerical results with the benchmark problem are shown and discussed in Section \ref{sec:ADRResults}.
Conclusions and future work are given in Section \ref{sec:Conclusion}.

\section{ADOL-C}\label{sec:adolc}

ADOL-C is an \emph{operator overloading} AD tool intended for differentiation of C programs
using C++ objects.
Other notable operator overloading tools include CoDiPack \cite{CoDiPack}, Sacado \cite{Sacado} and dco \cite{dco}.
Based on how chain rule is applied in AD tools, there are generally two differen modes, namely \emph{forward mode} and \emph{reverse mode}.

Consider a program \texttt{P} which evaluates a mathematical function $F:A\subset\mathbb R^n\rightarrow\mathbb R^m$.
By decomposing $F$ into a sequence of elementary operations, the \emph{forward mode} of AD differentiates \texttt{P} by repeated application of the \emph{chain rule}.
Propagating a \emph{seed vector} $\dot x\in\mathbb R^n$ through the differentiated program yields the Jacobian vector product (see (\ref{eq:ForwardReverse}));
appropriate selection of seed vectors (i.e. \emph{seed matrix}) enables computation of the Jacobian matrix itself. The \emph{reverse mode} of AD, on the other hand, propagates a seed vector $\bar y\in\mathbb R^m$ \emph{backwards}, yielding the Jacobian transpose vector product.
\begin{equation}\label{eq:ForwardReverse}
    \mbox{Forward mode:}\quad\dot y := \nabla F\:\dot x\enspace.\quad\mbox{Reverse mode:}\quad\bar x := (\nabla F)^T\bar y\enspace.
\end{equation}

Application of ADOL-C to an existing program requires identifying a section of the code to be differentiated and its marking as an \emph{`active'} section.
The user changes independent, dependent and intermediary variables of type \texttt{double} within the active section to be variables of the special ADOL-C type, \texttt{adouble}. That is, these variables are marked as \emph{`active variables'}.
Operations performed on active variables are recorded to a \emph{tape}, which is to be read again by ADOL-C.
Upon reading this tape for some input vector (of the same dimension as the number of independent variables), ADOL-C computes derivatives by overloading the operations recorded on the tape with their associated derivatives and invoking the chain rule appropriately.

\section{PETSc Adjoint}\label{sec:PETScAdjoint}

The adjoint approach used by PETSc works in a similar way as the reverse mode of AD, with the main difference being that the primitive operation becomes a time step instead of a line of source code.
A comprehensive reference to the discrete adjoint approach in PETSc can found in \cite{Zhang2017}.
Without loss of generality, we denote a multistage time stepping algorithm (e.g. Runge-Kutta methods) as 
\begin{equation}\label{eq:TSNumerical}
x_{n+1} = \N_n(x_n), \quad n=0,\dots,N-1,\quad x_0=\I,
\end{equation}
where $\I$ are the initial values and the solution at the end of the simulation is given by $x_N$.
The normal goal is to compute sensitivities of an objective function $\Psi = \psi (x_N;p)$
with respect to initial values or system parameters $p$.

By initializing the adjoint variable $\lambda$
with
\begin{equation}
 \lambda_{N} = \left( \frac{d \psi}{dx}(x_N) \right)^T
 \end{equation}
and solving the discrete adjoint model
\begin{equation}
 \lambda_{n}  =  \left( \frac{d \N}{d x} (x_n)  \right)^T \lambda_{n+1},\quad n= N-1, \dots, 0,
\label{eqn:disadjoint}
\end{equation}
backward in time, we obtain the gradient
$
\nabla_{\I} \psi(x_N) =  \lambda_0 .
$
Taking the simplest implicit method, backward Euler, for example.
The forward propagation equation for an autonomous ODE system is 
\begin{equation}
    x_{n+1} = x_n + h_n \, \f(x_{n+1})
    \label{eq:beuler}
\end{equation}
where $h_n$ is the stepsize at the $n$-th time step, and $\f$ is the ODE right-hand side function. 
The associated adjoint model will be
\begin{equation}
    \lambda_{n} = h_n \nabla_x\f (x_{n+1})^T \, \lambda_{n+1}.
\end{equation}

It can be seen that the essential ingredient in the adjoint computation is the Jacobian transpose.
When solving the original PDE problem, users need to supply Jacobian functions for solving the nonlinear system arising at each time step or use the finite-difference approximations constructed by PETSc; these Jacobians are reused in the adjoint computation as much as possible.
Extra Jacobians may be required by the adjoint solver depending on the problem settings (see \cite{Zhang2017}).
The PETSc adjoint solver assembles the adjoint calculation with the Jacobians at the timestepping level, thus avoiding a full differentiation of the whole library code, which consists of complicated data structure, iterative linear solvers and parallel infrastructure. However, it can still be a significant burden for users to provide callback functions to efficiently evaluate the Jacobians, especially when nonlinear PDE problems with complicated boundary conditions are involved.

\section{Benchmark Problem}\label{sec:example}

For the purposes of numerical experimentation and illustration of the automatic differentiation process we now introduce a nonlinear, time-dependent PDE.

The \emph{Gray-Scott model} describes the diffusion and reaction of two chemical species with concentrations $u$ and $v$, as in \cite{Gray1983}. The model may be expressed as a coupled system of scalar, nonlinear, time-dependent PDEs in two spatial dimensions \cite{Hundsdorfer2003}:
\begin{equation}\label{eq:GrayScott}
	\frac{\partial u}{\partial t} = D_1\Delta u-uv^2+\gamma(1-u)\enspace,\quad
	\frac{\partial v}{\partial t} = D_2\Delta v+uv^2-(\gamma+\kappa)v\enspace.
\end{equation}
Here $D_1,D_2>0$ and $\gamma,\kappa>0$ are diffusion and reactivity constants, respectively.

Consider a doubly periodic, square, spatial domain $\Omega=[0,2.5]\times[0,2.5]$ and a time interval $\mathcal T=(0,50]$.
As on (\cite{Hundsdorfer2003}, pp. 21-22), consider initial conditions
\begin{eqnarray}\label{eq:GrayScottIC1}
	u(x,y,0)=1-2v(x,y,0)\enspace,\\
\label{eq:GrayScottIC2}
	v(x,y,0)=\frac14\mathbbm{1}_{[1,1.5]}(x)\mathbbm{1}_{[1,1.5]}(y)\sin^2(4\pi x)\sin^2(4\pi y)\enspace.
\end{eqnarray}
For the adjoint analysis, we seek the sensitivity of the final solution at the domain centre with respect to
the initial conditions given by (\ref{eq:GrayScottIC1})--(\ref{eq:GrayScottIC2}).
It is essential to note that, given the nonlinear nature of the problem, the adjoint problem is directly dependent on the forward variables $(u,v)$ at every time step. 

A numerical solution is obtained using a centred finite difference discretization on a uniform quadrilateral mesh and Crank-Nicolson time integration \cite{Crank1947}.

One could argue that, due to the simplicity of Equation (\ref{eq:GrayScott}), the Jacobian may be easily derived by hand, rendering
% application of
AD unnecessary.
However, since we intend to construct a fully automated framework, we
take advantage of this opportunity and
derive also the analytic Jacobian to serve as a verification tool.

\section{Automating Jacobian Computation in PETSc}\label{sec:OutlineAndFeatures}

\subsection{Differentiated Program}\label{subsec:AD}

We consider an application of ADOL-C to generate the Jacobian of the discretized PDE residual \emph{at each timestep}.
The discretized residual code used by PETSc in solving the Gray-Scott model (\ref{eq:GrayScott}) is given in Listing \ref{lst:Passive}.
Given an array of \texttt{Field} structs containing the independent variables, $u$ and $v$, at some timestep, this function applies the spatial finite difference approximation in order to give the associated residual.

In the PETSc \texttt{TS} solver, data is stored using higher level structures such as matrices (\texttt{Mat}) and vectors (\texttt{Vec}).
Data may be accessed from these objects as arrays through the \texttt{DM} data management framework. In the case of Equation (\ref{eq:GrayScott}), we have a uniform, structured grid and so use the distributed array, \texttt{DMDA}.
In Listing \ref{lst:Passive}, arrays of \texttt{PetscScalar}s (i.e. \texttt{double}s) are extracted from the \texttt{DMDALocalInfo} and modified according to the residual evaluation.

From the user perspective, all that is required in order to apply AD to an existing PETSc code is to establish an \emph{`active version'} of the residual, wherein \emph{active variables} are defined and marked as independent or dependent, as appropriate.
Consider now Listing \ref{lst:Active}, which illustrates the active version of the code given in Listing \ref{lst:Passive}.
The calls to \texttt{trace\_on} and \texttt{trace\_off} indicate the active section of the code, which is to be differentiated and is identified by the global integer \texttt{tag}. Marking of the independence and dependence of variables occurs at the beginning and end of the active section, respectively.\footnote{The use of dummy variables in the marking of dependent variables corresponds to ghost points. Computations involving ghost points are required for parallelism, but these points should not be marked as dependent, as this would contribute additional nonzeros to the Jacobian.} In between, the loops which evaluate the residual take exactly the same form as in Listing \ref{lst:Passive}, except that the \texttt{double} variables involved in the residual evaluation have been replaced by active variables of \texttt{adouble} type.

For each instance in which the Jacobian is to be evaluated, ADOL-C overloads operations on the tape and propagates seed vectors through the forward (or reverse) mode of AD, thereby computing derivatives. In the case of (\ref{eq:GrayScott}), the same tape can be used for all timesteps, since the form of the discretized residual does not explicitly depend on the time level.
As such, the active version of the residual is evaluated once, \emph{before} the time integration, with the original, \emph{`passive'} version used thereafter.
This means the computational cost can be reduced by only generating the tape once. In the general case, the `active' version of the residual is used at every function evaluation.

In the following two subsections, we investigate effeciency considerations for the Jacobian computation using ADOL-C. In the implementations of both Subsection \ref{subsec:MatrixAssembly} and \ref{subsec:MatrixFree}, Listing \ref{lst:Active} is used to generate the tape. The difference between these methods is in how the tape is used, not how it is generated.

\begin{figure}[t]
\begin{lstlisting}[label={lst:Passive},caption={Passive residual evaluation for Equation (\ref{eq:GrayScott}).},basicstyle=\scriptsize\ttfamily]
PetscErrorCode IFunctionLocalPassive(DMDALocalInfo *info,PetscReal t,Field**u,Field**udot,Field**f,void *ptr)
{
  AppCtx         *appctx = (AppCtx*)ptr;
  PetscInt       i,j,xs,ys,xm,ym;
  PetscReal      hx,hy,sx,sy;
  PetscScalar    uc,uxx,uyy,vc,vxx,vyy;

  PetscFunctionBegin;
  hx = 2.50/(PetscReal)(info->mx); sx = 1.0/(hx*hx);
  hy = 2.50/(PetscReal)(info->my); sy = 1.0/(hy*hy);
  /* Get local grid boundaries */
  xs = info->xs; xm = info->xm; ys = info->ys; ym = info->ym;
  /* Compute function over the locally owned part of the grid */
  for (j=ys; j<ys+ym; j++) {
    for (i=xs; i<xs+xm; i++) {
      uc        = u[j][i].u;
      uxx       = (-2.0*uc + u[j][i-1].u + u[j][i+1].u)*sx;
      uyy       = (-2.0*uc + u[j-1][i].u + u[j+1][i].u)*sy;
      vc        = u[j][i].v;
      vxx       = (-2.0*vc + u[j][i-1].v + u[j][i+1].v)*sx;
      vyy       = (-2.0*vc + u[j-1][i].v + u[j+1][i].v)*sy;
      f[j][i].u = udot[j][i].u - appctx->D1*(uxx + uyy) + uc*vc*vc - appctx->gamma*(1.0 - uc);
      f[j][i].v = udot[j][i].v - appctx->D2*(vxx + vyy) - uc*vc*vc + (appctx->gamma + appctx->kappa)*vc;
    }
  }
  PetscFunctionReturn(0);
}
\end{lstlisting}
\end{figure}

\begin{figure}[p]
\begin{lstlisting}[label={lst:Active},caption={Active residual evaluation for Equation (\ref{eq:GrayScott}).},basicstyle=\scriptsize\ttfamily]
PetscErrorCode IFunctionLocalActive(DMDALocalInfo *info,PetscReal t,Field**u,Field**udot,Field**f,void *ptr)
{
  AppCtx         *appctx = (AppCtx*)ptr;
  AField         **f_a = appctx->f_a,**u_a = appctx->u_a;
  PetscInt       i,j,xs,ys,xm,ym,gxs,gys,gxm,gym;
  PetscReal      hx,hy,sx,sy;
  PetscScalar    dummy;
  adouble        uc,uxx,uyy,vc,vxx,vyy;

  PetscFunctionBegin;
  hx = 2.50/(PetscReal)(info->mx); sx = 1.0/(hx*hx);
  hy = 2.50/(PetscReal)(info->my); sy = 1.0/(hy*hy);
  xs = info->xs; xm = info->xm; gxs = info->gxs; gxm = info->gxm;
  ys = info->ys; ym = info->ym; gys = info->gys; gym = info->gym;
  trace_on(tag);
  /* Mark independence */
  for (j=gys; j<gys+gym; j++) {
    for (i=gxs; i<gxs+gxm; i++) {
      u_a[j][i].u <<= u[j][i].u;u_a[j][i].v <<= u[j][i].v;
    }
  }
  /* Compute function over the locally owned part of the grid */
  for (j=ys; j<ys+ym; j++) {
    for (i=xs; i<xs+xm; i++) {
      uc          = u_a[j][i].u;
      uxx         = (-2.0*uc + u_a[j][i-1].u + u_a[j][i+1].u)*sx;
      uyy         = (-2.0*uc + u_a[j-1][i].u + u_a[j+1][i].u)*sy;
      vc          = u_a[j][i].v;
      vxx         = (-2.0*vc + u_a[j][i-1].v + u_a[j][i+1].v)*sx;
      vyy         = (-2.0*vc + u_a[j-1][i].v + u_a[j+1][i].v)*sy;
      f_a[j][i].u = udot[j][i].u - appctx->D1*(uxx + uyy) + uc*vc*vc - appctx->gamma*(1.0 - uc);
      f_a[j][i].v = udot[j][i].v - appctx->D2*(vxx + vyy) - uc*vc*vc + (appctx->gamma + appctx->kappa)*vc;
    }
  }
  /* Mark dependence */
  for (j=gys; j<gys+gym; j++) {
    for (i=gxs; i<gxs+gxm; i++) {
      if ((i < xs) || (i >= xs+xm) || (j < ys) || (j >= ys+ym)) {
        f_a[j][i].u >>= dummy;f_a[j][i].v >>= dummy;
      } else {
        f_a[j][i].u >>= f[j][i].u;f_a[j][i].v >>= f[j][i].v;
      }
    }
  }
  trace_off();
  PetscFunctionReturn(0);
}
\end{lstlisting}
\end{figure}

\subsection{Jacobian Assembly using ADOL-C}\label{subsec:MatrixAssembly}

Let $n$ be the number of \emph{degrees of freedom (DOFs)} at each timestep of the discretization of problem (\ref{eq:GrayScott}).
Assembling the Jacobian $J\in\mathbb R^{n\times n}$ using ADOL-C, requires a \emph{seed matrix} $\mathcal S\in\mathbb R^{n\times p}$, where $p\leq n$.
The `natural' approach is to choose the identity matrix, $\mathcal S=\mathcal I_n$
(whence $p=n$). Propagating the entire identity matrix through the forward mode of AD is typically both prohibitively expensive and unnecessary, as illustrated in Section \ref{sec:ADRResults}.
Selecting a seed matrix with second dimension $p\ll n$ would induce great computational savings.

If the Jacobian sparsity pattern, $\mathcal P$, is known, reductions in $p$ can be achieved by \emph{coloring} \cite{ColPack}:
columns of the Jacobian may be assigned the same \emph{color} if there exist no rows where both have nonzero entries.
By this color identification, we are able to \emph{losslessly} compress the Jacobian, reducing its size by $n-p$ columns.
A seed matrix, $\mathcal S$, is chosen with columns corresponding to the colors used, entries of unity where the row index corresponds to an accordingly colored column in the Jacobian and zeroes elsewhere.
Propagating $\mathcal S$ through the forward mode of AD yields a \emph{compressed} Jacobian, $J_C\in\mathbb R^{n\times p}$.
This equates to propagating $p$, as opposed to $n$, vectors, thereby reducing the cost of the propagation.
Computational savings induced by this compressed approach
are inherently problem and discretization dependent: the smaller the stencil, the smaller $p$ can be chosen.

The Jacobian can be recovered by lossless de-compression, as first observed by \cite{Averick1994}.
Non-negative entries of the resulting `recovery matrix', $\mathcal R\in\mathbb R^{n\times p}$, hold column indices for the corresponding entries in $J_C$, whilst negative entries correspond to zeroes.
For $p\ll n$, the additional computational cost associated with applying the index mapping of $\mathcal R$ to assemble $J$ is assumed to be much less than the savings of the propagation step.

Algorithm \ref{alg:Compressed} outlines the compressed Jacobian computation procedure.
The first step is performed by the graph coloring package, ColPack \cite{ColPack}.
Given the boundary conditions and stencil used, PETSc is able to generate a coloring for the second step.
Forward propagation of $\mathcal S$ in the third step is performed by ADOL-C, using what was written to tape.
The final step amounts to de-compressing $J_C$ to give the Jacobian, by application of the index mapping associated with $\mathcal R$.

\begin{algorithm}[h!]
    Extract the sparsity pattern $\mathcal P$ of the Jacobian from the tape\\
    Establish a coloring from $\mathcal P$ and thereby generate a \emph{seed matrix} $\mathcal S$\\
    Compute the compressed Jacobian, using $\mathcal S$ and the tape\\
    Recover the Jacobian from the compressed format using $\mathcal P$ and $\mathcal S$\\
	\caption{Four step approach to sparse Jacobian computation \cite{ColPack}.}
	\label{alg:Compressed}
\end{algorithm}

Jacobian evaluation for the approach described in this subsection is achieved in this work using a problem-independent PETSc-ADOL-C utility function which performs the required propagation and recovery steps and assembles the associated \texttt{Mat} object.

\subsection{Jacobian-Free Implementation using ADOL-C}\label{subsec:MatrixFree}

For large problems, (compressed or uncompressed) Jacobian assembly, as described in Subsection \ref{subsec:MatrixAssembly}, can become memory intensive.
Furthermore, the Jacobian may only be used a few times, whereby its assembly might be seen as wasteful of computational resources.

In the numerical solution of time-dependent, nonlinear PDEs, a nonlinear solve is performed at each timestep, which in turn relies on the solution of linear systems involving the Jacobian.
For Krylov subspace methods, the Jacobian is not explicitly needed
- only its \emph{action} upon a vector.
This motivates \emph{matrix-free} methods, wherein the seed vector is chosen as the vector the Jacobian acts upon.
For the adjoint solve, the Jacobian transpose vector product can be provided by the reverse mode of AD.

The implementation of the matrix-free Jacobian evaluation differs from that described in Subsection \ref{subsec:MatrixAssembly}.
The PETSc \texttt{MATSHELL} matrix type is used to endow the \texttt{Mat} object with additional context information, whereupon we may overload the \text{MatMult} operations associated with Jacobian vector products so that they apply the forward mode of AD directly to the given input vector. Similarly, the \text{MatMultTranspose} operations associated with Jacobian-transpose vector products are overloaded so as to directly apply the reverse mode of AD.

\section{Experimental Results}\label{sec:ADRResults}

This section compares the strategies discussed above using performance analysis.
Due to the non-triviality of effectively preconditioning an AD-generated matrix-free algorithm, preconditioning has not been used in any of the following results. The adjoint problem is solved using checkpointing to memory.

\begin{figure}[b!]
    \centering
    \begin{subfigure}{0.315\textwidth}
        \centering
        \includegraphics[width=0.95\textwidth]{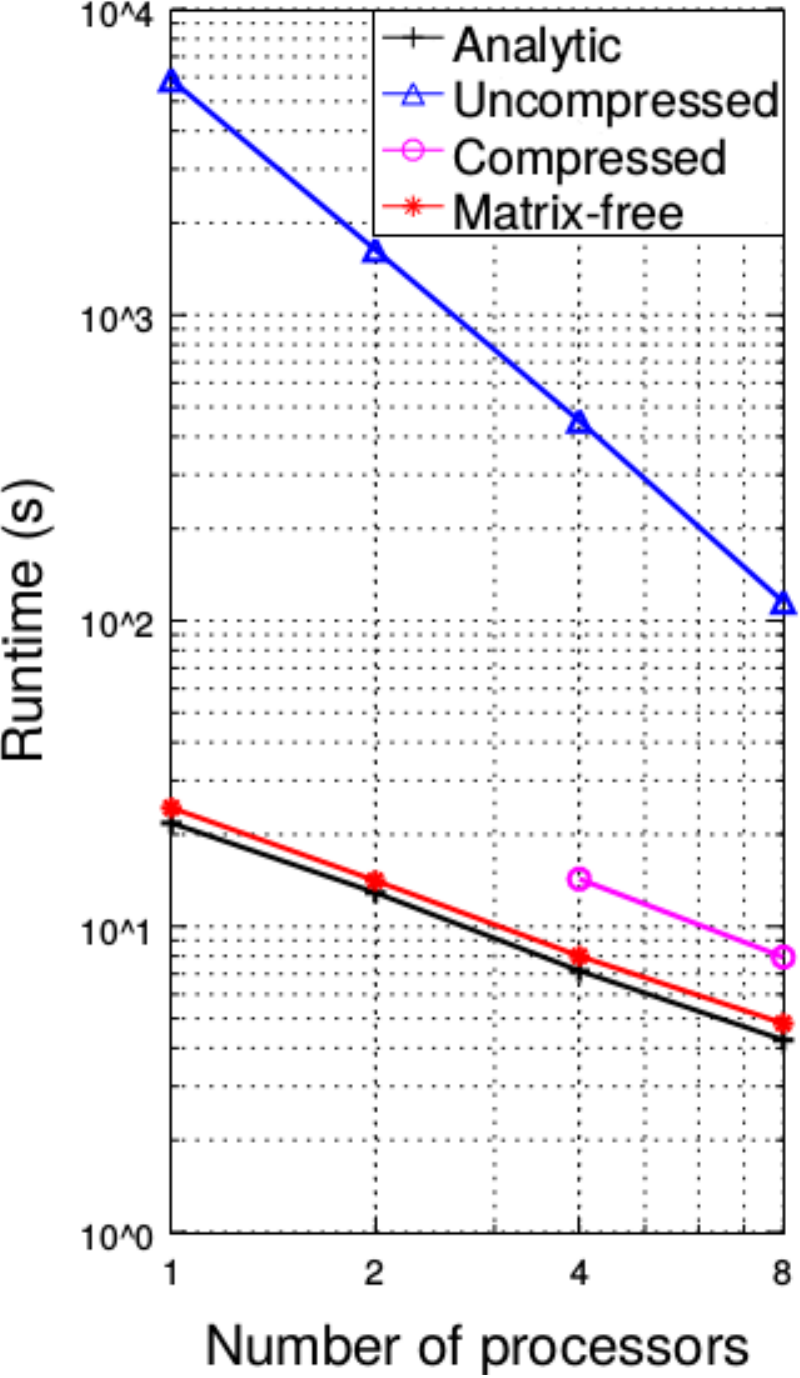}
        \caption{$65\times65$}
        \label{fig:runtimes65}
    \end{subfigure}
    \,
    \begin{subfigure}{0.315\textwidth}
        \centering
        \includegraphics[width=0.95\textwidth]{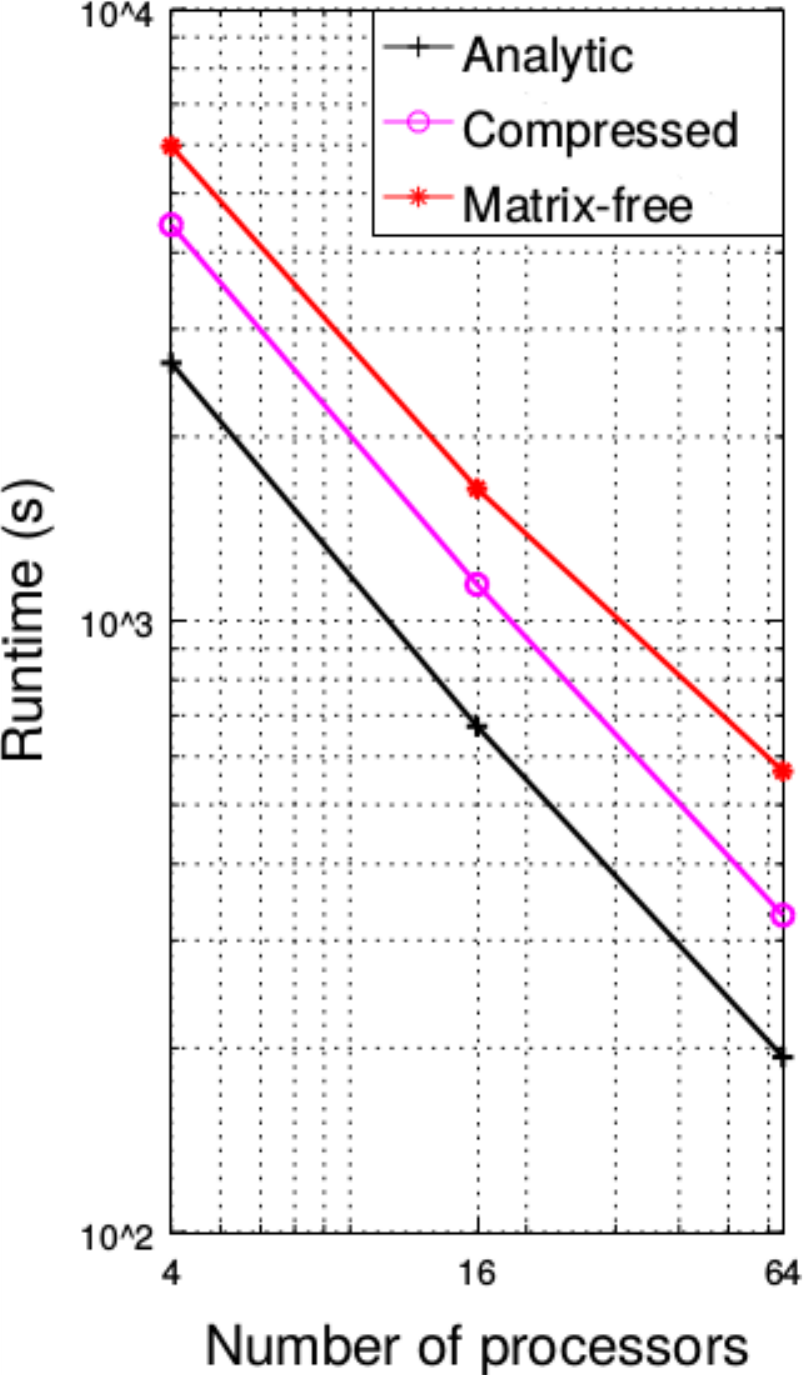}
        \caption{$1000\times1000$}
        \label{fig:runtimes1000}
    \end{subfigure}
    \,
    \begin{subfigure}{0.315\textwidth}
        \centering
        \includegraphics[width=0.95\textwidth]{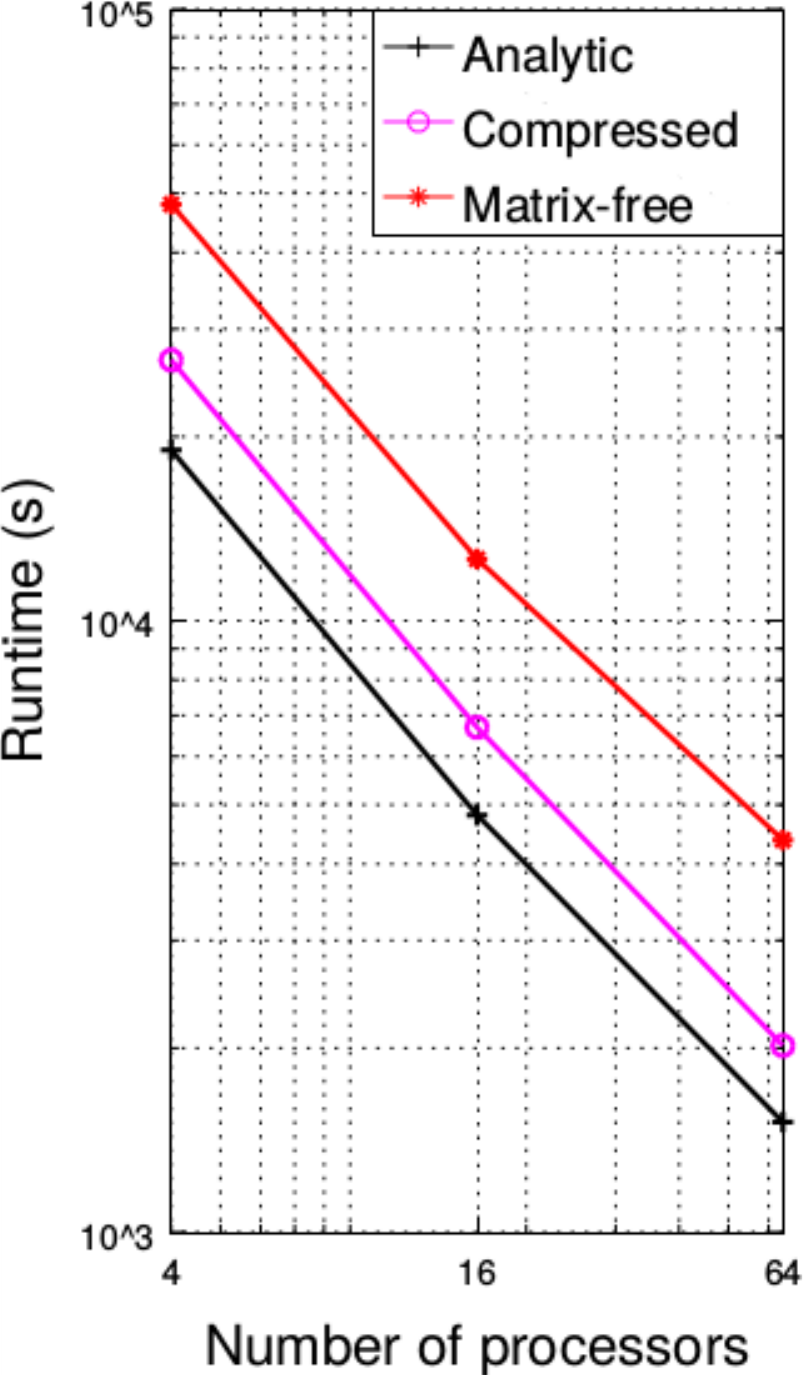}
        \caption{$2000\times2000$}
        \label{fig:runtimes2000}
    \end{subfigure}
    \caption{
    Total runtimes for given grid resolutions.
    Time discretisation: 100 timesteps, $\Delta t=0.5$.
%    Architecture: Intel Xeon Phi Knights Landing.
    }
    \label{fig:runtimes}
\end{figure}

Figure \ref{fig:runtimes65} illustrates that, even for a $65\times65$ grid, using ADOL-C to propagate the entire identity matrix through the forward mode of AD (i.e. \emph{`uncompressed'} Jacobian assembly) is prohibitively expensive, although performance is greatly increased by parallelism.
For higher resolutions, the approach is completely infeasible, since the number of active variables becomes too large.
As such, it is imperative that the approaches of Subsections \ref{subsec:MatrixAssembly}--\ref{subsec:MatrixFree} are considered.

\begin{figure}[t!]
    \centering
    \hfill
    \begin{subfigure}[t]{0.2\textwidth}
        \centering
        \includegraphics[width=0.99\textwidth]{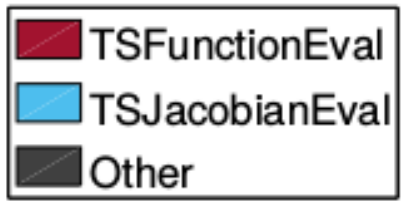}
    \end{subfigure}
    \\
    \begin{subfigure}{0.31\textwidth}
        \centering
        \includegraphics[width=0.99\textwidth]{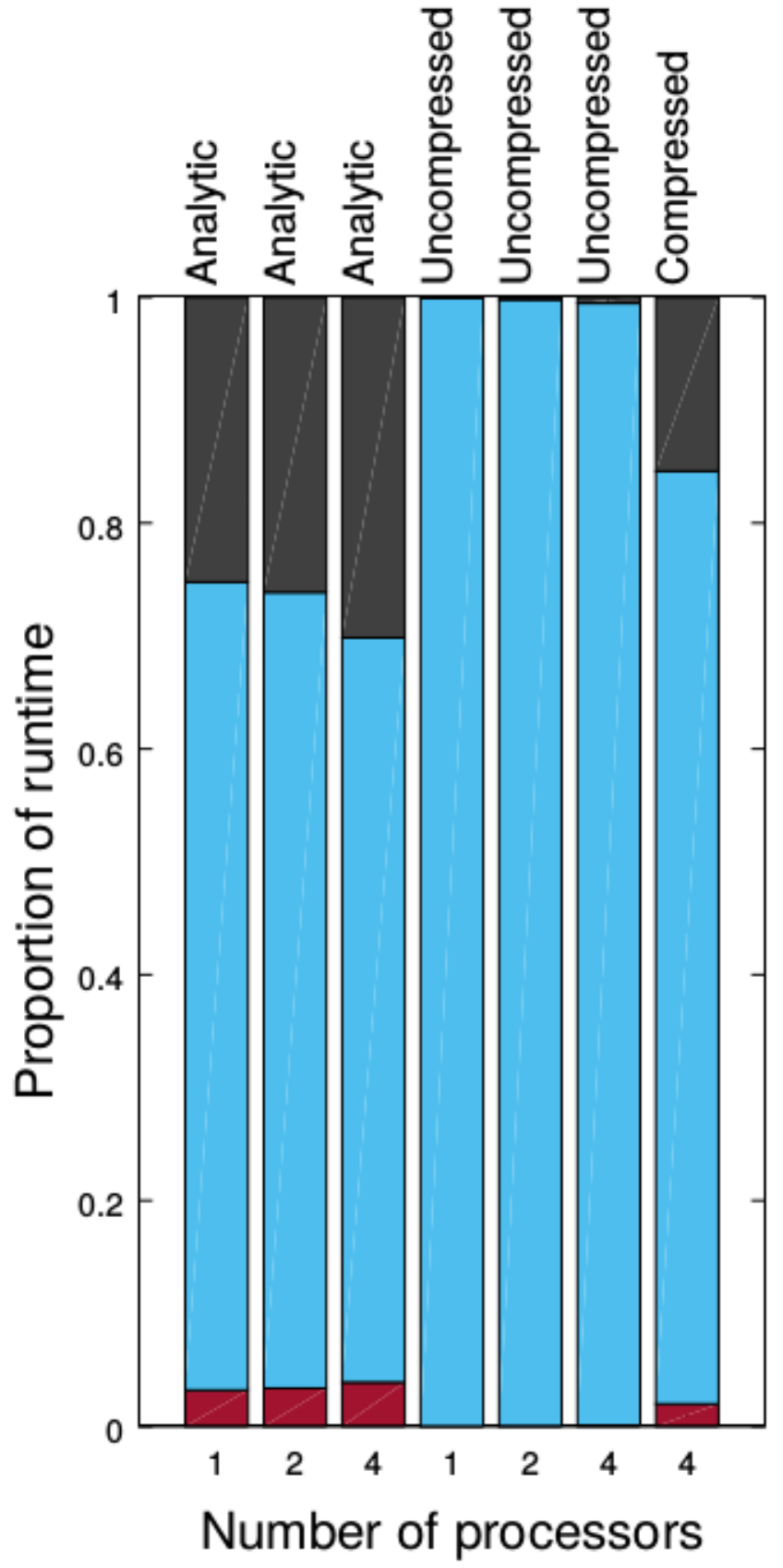}
        \caption{$65\times65$ grid}
        \label{fig:componentsTS65}
    \end{subfigure}
    \,
    \begin{subfigure}{0.31\textwidth}
        \centering
        \includegraphics[width=0.99\textwidth]{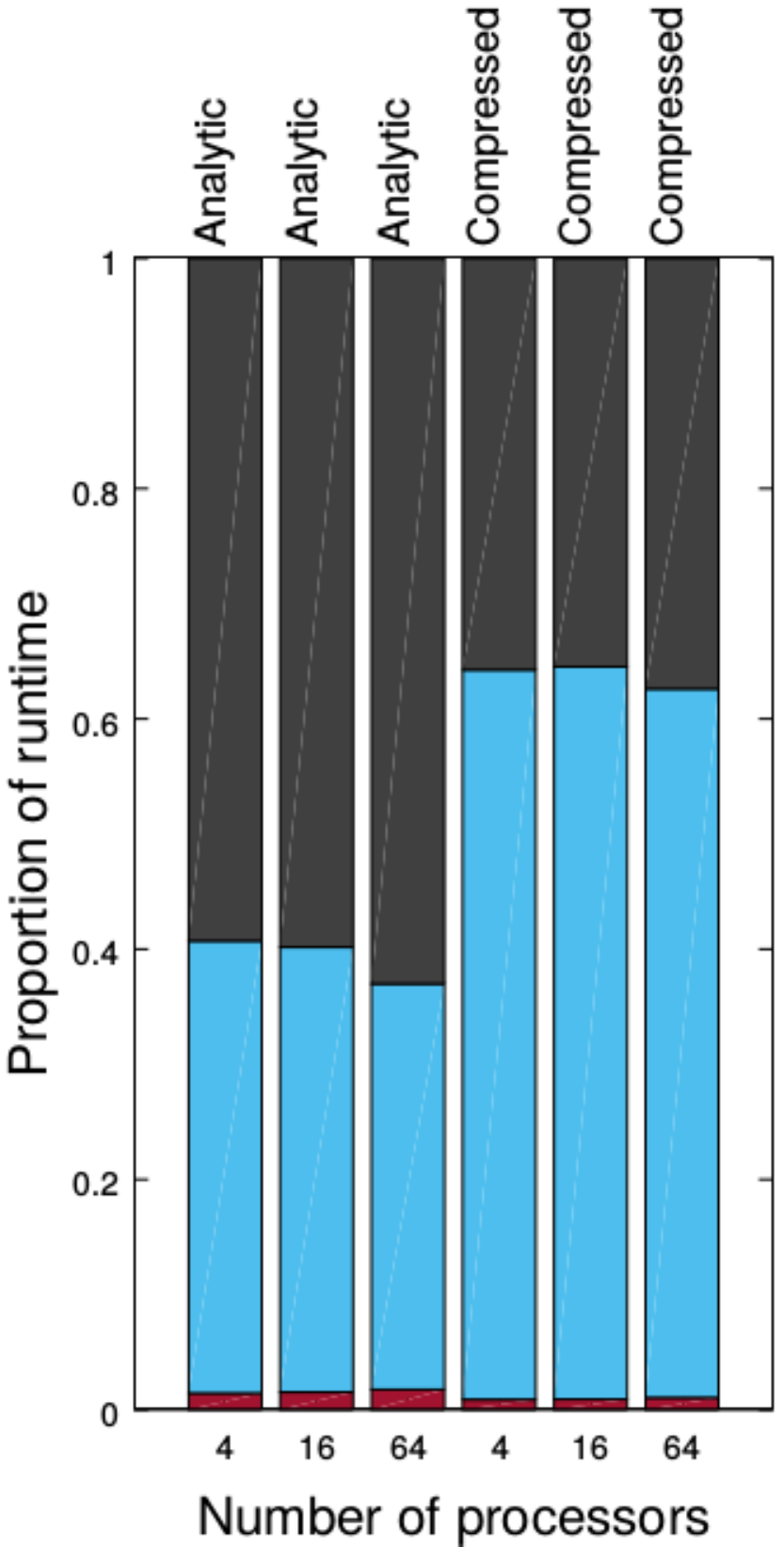}
        \caption{$1000\times1000$ grid}
        \label{fig:componentsTS1000}
    \end{subfigure}
    \,
    \begin{subfigure}{0.31\textwidth}
        \centering
        \includegraphics[width=0.99\textwidth]{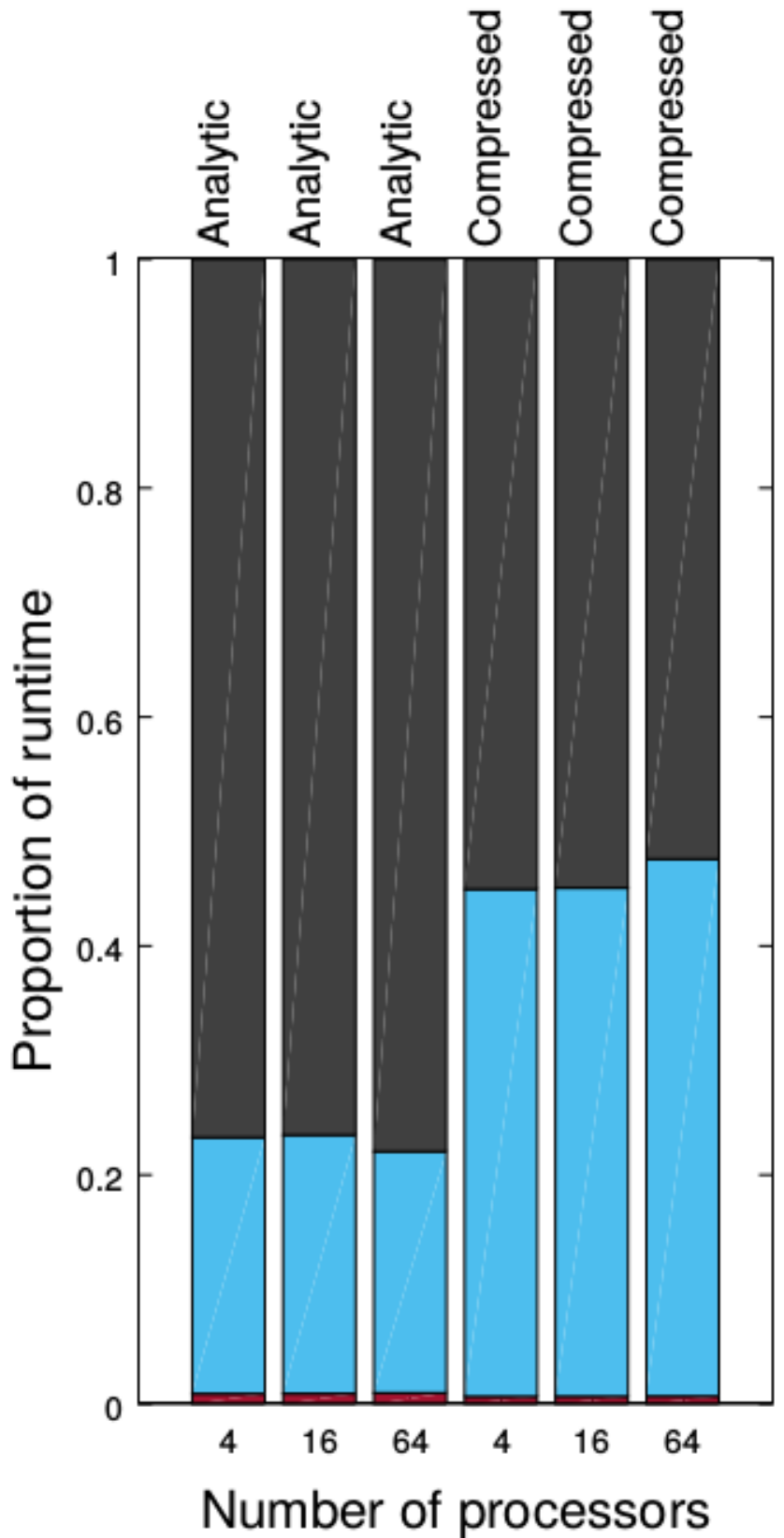}
        \caption{$2000\times2000$ grid}
        \label{fig:componentsTS2000}
    \end{subfigure}
    \caption{Time integration runtimes for the three assembled Jacobian techniques considered. `Analytic' corresponds to the case in which the Jacobian is derived analytically, whilst `uncompressed' corresponds to the case where the Jacobian is computed by passing the identity matrix through the forward mode of AD.}
    \label{fig:componentsTS}
\end{figure}

Applying coloring techniques can significantly reduce this cost. In Fig. \ref{fig:runtimes}, the compressed Jacobian approach exhibits strong scalability.
%\footnote{There are no results on one or two processors because of the double periodicity of the domain. The PETSc coloring strategy (\texttt{DM\_COLORING\_LOCAL}) used does not allow any processor to own nodes which are identified by the boundary condition.}
The runtime of this approach is less than twice that of the hand-coded approach and this relative performance actually improves as problem size increases.
Since the assembled Jacobian approaches have the same workflow, we may directly compare runtimes for their constituent operations. Figure \ref{fig:componentsTS} highlights Jacobian evaluation as the computational bottleneck, particularly in cases where AD is applied.
As such, it is important that the Jacobian is computed as efficiently as possible.

As revealed in Subsection \ref{subsec:MatrixAssembly}, there are just two differences between the analytic and compressed approaches. Firstly, in the latter, the sparsity pattern, seed matrix and recovery matrix must be established before the time integration begins. Secondly, whenever the Jacobian is to be assembled, the seed matrix is to be propagated through the forward mode of AD and then decompressed. As such, the discrepancies in absolute runtime illustrated in Fig. \ref{fig:runtimes} and proportions of runtime spent in Jacobian evaluation illustrated in Fig. \ref{fig:componentsTS} may be explained as being due to the AD propagation and compression algorithm.

% \hong{It is great that the gap shrinks when the problem size increased. But isn't the compressed Jacobian approach supposed to be really close to the Analytic version? We need to explain why the little bit performance is lost. Does the compressed Jacobian generated by ADOL-c have exactly the same sparity pattern as the original Jacobian? Compared to the analytical version, what is the extra work?}

\begin{figure}[t]
    \centering
    \hfill
    \begin{subfigure}[t]{0.2\textwidth}
        \centering
        \includegraphics[width=0.99\textwidth]{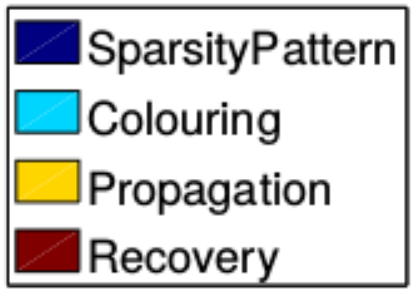}
    \end{subfigure}
    \\
    \begin{subfigure}{0.48\textwidth}
        \centering
        \includegraphics[width=0.99\textwidth]{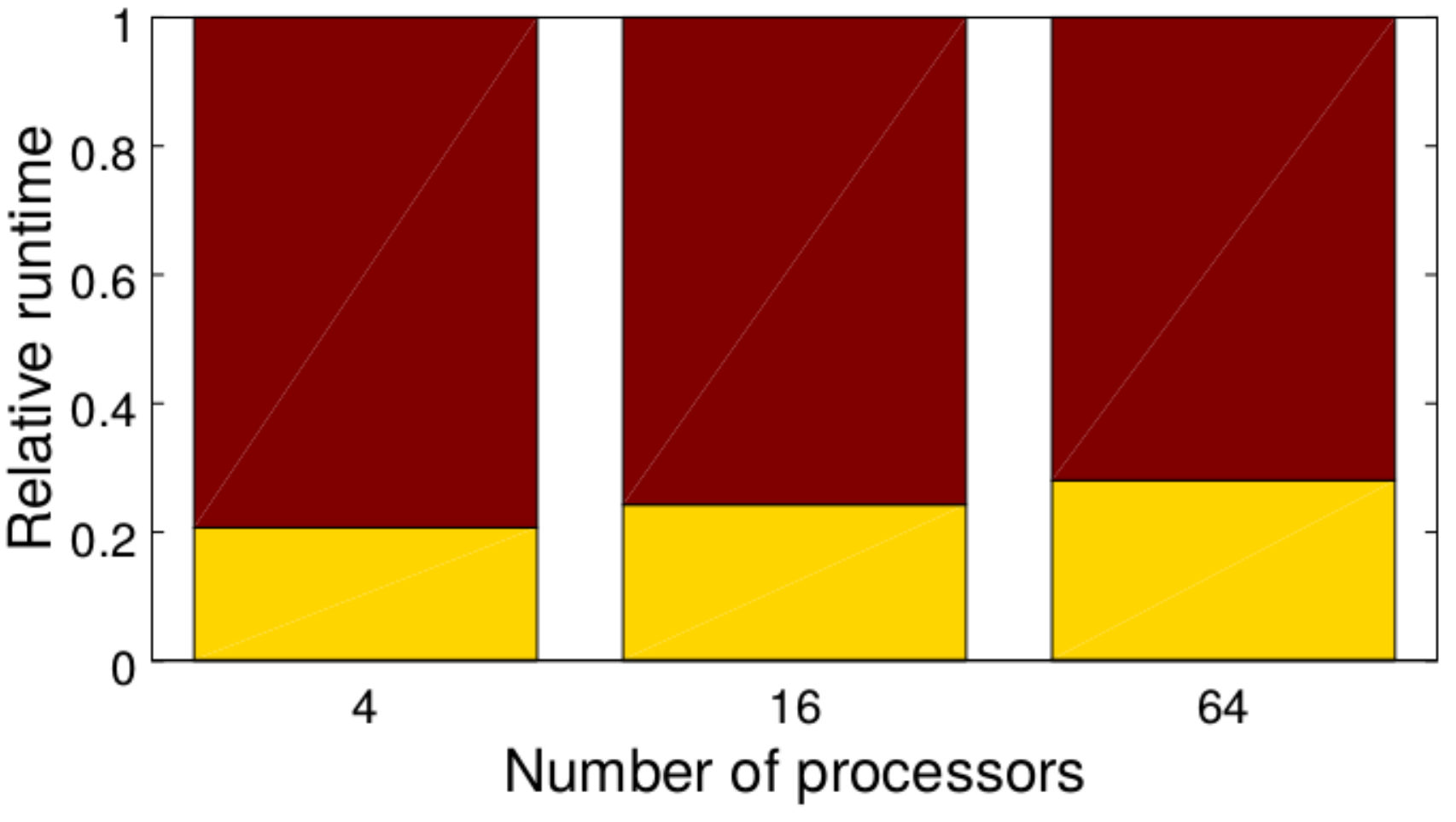}
        \caption{$1000\times1000$ grid}
        \label{fig:compressed1000}
    \end{subfigure}
    \,
    \begin{subfigure}{0.48\textwidth}
        \centering
        \includegraphics[width=0.99\textwidth]{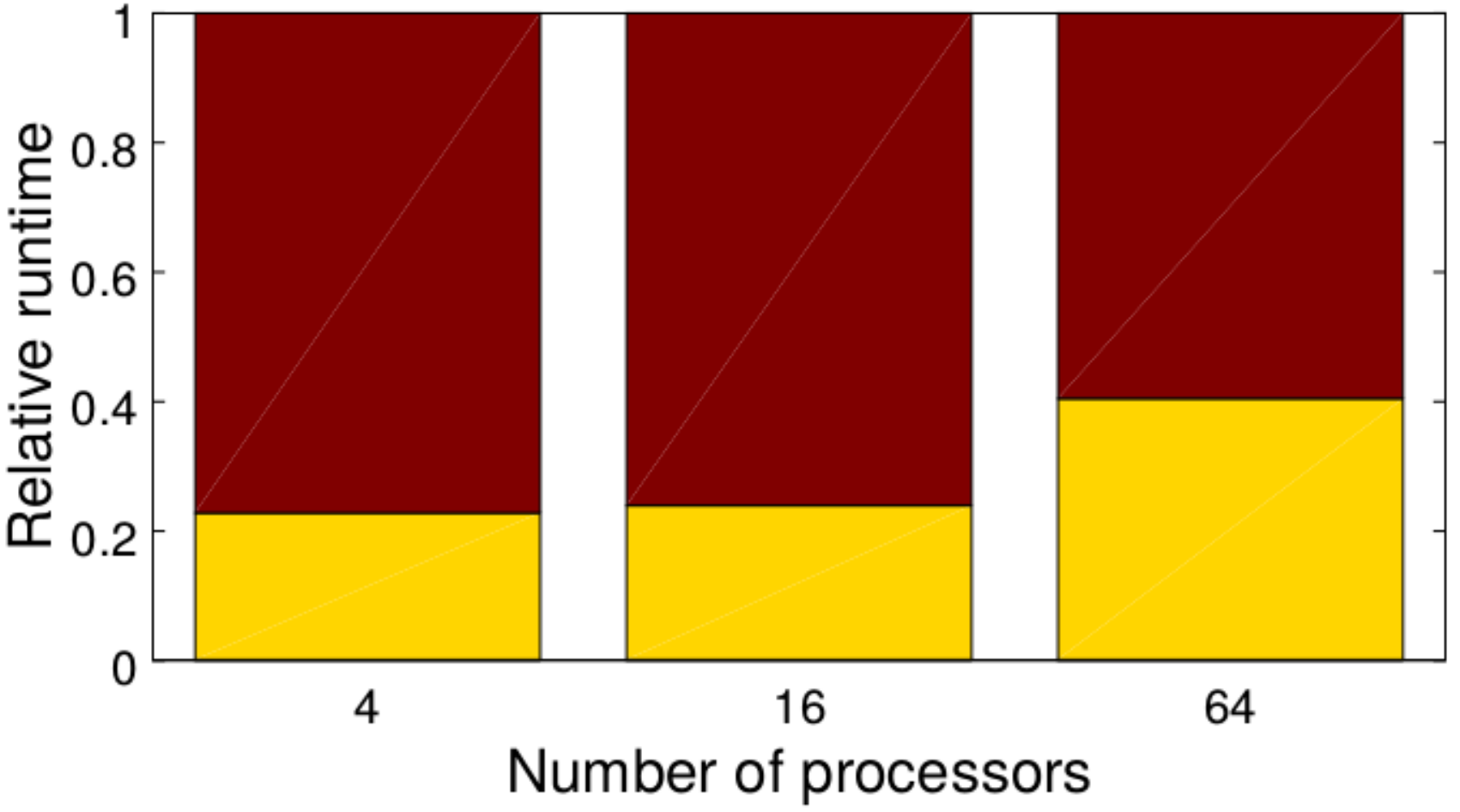}
        \caption{$2000\times2000$ grid}
        \label{fig:compressed2000}
    \end{subfigure}
    \caption{Comparative runtimes for steps of Algorithm \ref{alg:Compressed}, for the given grid resolutions.}
    \label{fig:compressed}
\end{figure}
Figure \ref{fig:compressed} further breaks down the runtimes of the compressed Jacobian evaluation steps, indicating that the cost of generating the sparsity pattern and seed matrix are negligible.
This is to be expected, since they are generated only once, whilst forward propagation and Jacobian recovery occur at every timestep.
We may deduce from Fig. \ref{fig:compressed} that the discrepancies between the analytic and compressed approaches shown in Fig. \ref{fig:runtimes} are almost entirely due to the propagation and recovery steps of Algorithm \ref{alg:Compressed}.
The simple recovery routine used for the final step is shown to be \emph{more costly} than the forward propagation, indicating potential inefficiencies in the simple recovery technique used.
The discrepancy between analytic and compressed approaches would be reduced through implementation of a more efficient recovery technique.
Nonetheless, the performance shown in Fig. \ref{fig:runtimes} illustrates that the inefficient recovery technique does not pose a serious problem in this particular application.

On the $65\times65$ grid, the matrix-free approach performs almost as well as assembling the analytic Jacobian.
For higher grid resolution, Subfigures \ref{fig:runtimes1000}--\ref{fig:runtimes2000} indicate that this relative performance diminishes somewhat.
Table \ref{tab:KSPits} show linear solver iterations growing with spatial resolution, indicating the problem becoming increasingly ill-conditioned.
% It is also noteworthy that the matrix-free algorithm generally requires fewer linear solver iterations than the assembled approaches.
% \hong{This is not shown anywhere, and there is no reason that matrix-free requires fewer iterations. It is mostly like an effect due to round-off errors as we discussed in the emails.}
Variations in the number of linear solver iterations displayed in each column of Table \ref{tab:KSPits} can be explained as being due to round-off error, the influence of which is essentially unpredictable.
Whilst Jacobian assembly occurs at each iteration of the \emph{nonlinear} solver, the matrix-free approach applies AD at each iteration of the \emph{linear} solver.
\begin{table}[t]
    \centering
    \begin{tabular}{|c|c|c||c|c|c|c|c|}
        \hline
         \multicolumn{3}{|c||}{$65\times65$}& \multicolumn{3}{|c|}{$1000\times1000$} & \multicolumn{2}{|c|}{$2000\times2000$}\\
        \hline
        Proc. & Assembled & Matrix-free & Proc. & Assembled & Matrix-free & Assembled & Matrix-free \\
        \hline
        1 & 1763 & 1763 & 4  & 8320 & 8113 & 17350 & 15953\\
        2 & 1763 & 1766 & 16 & 8322 & 8165 & 17255 & 15960\\
        4 & 1763 & 1762 & 64 & 8323 & 8167 & 17251 & 16053\\
        8 & 1763 & 1762 &    &      &      &       & \\
        \hline
    \end{tabular}
    \caption{Linear solver iterations as a function of grid resolution and number of processors.}
    \label{tab:KSPits}
\end{table}

For the compressed approach and for each grid resolution, there are two nonlinear solver iterations at each timestep, each involving the forward propagation of five vectors (corresponding to the five colors used). This gives ten forward propagations per timestep.
For the matrix-free approach, on the other hand, a single vector is propagated forwards whenever PETSc's \texttt{MatMult} function is invoked, corresponding to the computation of a Jacobian-vector product. This occurs, on average, 36 and 76 times per timestep, for the $1000\times1000$ and $2000\times2000$ grids, respectively. Similar logic can be applied to the reverse propagations in the adjoint steps.

Given that the application of AD is the computational bottleneck, these remarks indicate why, despite the matrix-free approach often using fewer Krylov iterations, runtimes are longer than for the compressed Jacobian.
A preconditioning strategy which ensures that the number of linear solves is independent of mesh resolution would help to alleviate this disadvantage.
A key observation is that the compressed approach is preferable for nonlinear problems wherein the Jacobian needs to be updated frequently.
Nevertheless, Figure \ref{fig:runtimes} shows that both compressed and matrix-free approaches have runtime of the same order as an analytic Jacobian and are therefore both valuable strategies for automatic Jacobian computation.
% \hong{We want to highlight the compressed approach is more preferrable for nonlinear problems where Jacobian needs to be updated frequently.}

\section{Conclusion}\label{sec:Conclusion}

We illustrated an application of ADOL-C to automatically provide Jacobian transforms for time-dependent, nonlinear PDEs and their adjoints, solved using PETSc. Whilst the Gray-Scott model (\ref{eq:GrayScott}) considered is fairly simple, the differentiation procedure may be reused for any complex problem.

Efficiency considerations are made, including compressed and matrix-free methods. These enable ADOL-C to automatically generate Jacobian transforms for larger problems, within a timeframe of the same order as the case where the Jacobian is given.
We illustrate the effectiveness of the compressed approach for a finite difference problem with a relatively small stencil. The compression and propagation algorithm used exhibits strong scalability for the numbers of processors and grid sizes considered.
Whilst the Jacobian-free approach performs promisingly at low resolution, it suffers when applied to larger, unconditioned problems and would greatly benefit from an effective preconditioning strategy.

In the case of Jacobian assembly using ADOL-C, the task of preconditioning is only as difficult as for an analytically derived Jacobian;
a wide variety of preconditioners are already implemented in PETSc for this purpose.
One downside of the matrix-free approach in PETSc is that we cannot use any of the built-in preconditioners, meaning users must provide their own, using the \texttt{PCSHELL} preconditioner type.
Furthermore, effective preconditioning for the transposed solve in the adjoint model is challenging and remains an open research problem.
% \hong{AFAIK, we dont have PC for the matrix-free case, do we? The downside of matrix-free here is we cannot use any of the built-in PC and users have to provide their custom PC for matrix-free . Furthermore PC for the transposed solve in the adjoint model is challenging and remains an open research problem}

Suppose we choose to implement Jacobi preconditioning - one of the simplest forms of preconditioning.
In a matrix-free approach, it is not possible to directly compute the action of the (inverse) Jacobian diagonal upon a seed vector using the same trace as for the Jacobian application.
Assembling the Jacobian diagonal alone using a compressed method requires the same number of seed vectors to be propagated through AD as would be required to assemble the Jacobian itself.
Assembling the diagonal in this way would of course defeats the point of using matrix-free.
However, one advantage of Jacobi preconditioning is that the preconditioner can be reused for the adjoint time integration.
Depending on the problem at hand, it may be possible to make a separate trace of the PDE residual, with different independent and dependent variables, in order to implement a matrix-free preconditioner with ADOL-C.
% Another possibility is to use a \emph{staggered} preconditioner, thereby avoiding Jacobian assembly at each timestep.
% However, there is a trade-off here between how often the Jacobian is to be assembled and the potential increase in linear iteration count as the preconditioner is reused and becomes 

Whilst Jacobian compression is more performant for the Gray-Scott model with low order centred finite differences, this would not necessarily be the case for PDEs and discretizations which have larger stencils and therefore require more colors.
Additionally, larger stencils would make the sub-optimal Jacobian recovery routine highlighted in Figure \ref{fig:compressed} more problematic.
Replacing this routine by one provided by ColPack would likely improve performance.

Note that the PDE residual code illustrated in Listings \ref{lst:Passive} and \ref{lst:Active} involve doubly nested loops which operate on the components of arrays directly.
In some applications, it is beneficial to avoid accessing these array components directly and instead perform operations on the higher level PETSc \texttt{Vec} or \texttt{Mat} structures.
Future work might also consider raising the abstraction level (see \cite{Bischof1996}), in order to differentiate such expressions, thereby simplifying the process of applying AD to complex PETSc codes.
When raising the abstraction level, further computational savings can be attained through awareness of the PDE discretization;
appropriate data abstraction allows the preservation of efficiency choices made in the problem discretization.

\subsection*{Acknowledgements}\label{sec:Acknowledgements}
Many thanks to Argonne's Autodiff research group and PETSc development team, particularly Sri Hari Krishna Narayanan, Michel Schanen, Jan H\"uckelheim, Barry Smith and Navjot Kukreja.

\bibliographystyle{splncs04}
\bibliography{paper}

\newpage
\thispagestyle{empty}
{\bf Government License.}  The submitted manuscript has been created by
UChicago Argonne, LLC, Operator of Argonne National Laboratory
(``Argonne''). Argonne, a U.S. Department of Energy Office of Science
laboratory, is operated under Contract No. DE-AC02-06CH11357. The
U.S. Government retains for itself, and others acting on its behalf, a
paid-up nonexclusive, irrevocable worldwide license in said article to
reproduce, prepare derivative works, distribute copies to the public,
and perform publicly and display publicly, by or on behalf of the
Government.  The Department of Energy will provide public access to
these results of federally sponsored research in accordance with the
DOE Public Access
Plan. http://energy.gov/downloads/doe-public-access-plan.
\end{document}